\documentclass{article}
\usepackage[T1]{fontenc} %
\usepackage[utf8]{inputenc} %
\usepackage{ismir,amsmath,cite,url}
\usepackage{graphicx}
\usepackage{color}
\usepackage{multirow}
\usepackage{microtype}
\usepackage{adjustbox}

\usepackage{caption}
\usepackage{subcaption}

\usepackage{lineno}

\title{On the Effectiveness of Speech Self-Supervised Learning for Music}

\multauthor
{\small Yinghao Ma \textsuperscript{\includegraphics[scale=0.03]{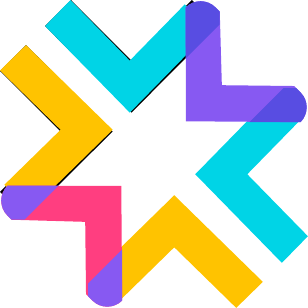},1*} \hspace{0.15cm} Ruibin Yuan \textsuperscript{\includegraphics[scale=0.03]{figs/map-logo-c.pdf},2,3*} \hspace{0.15cm} Yizhi Li \textsuperscript{\includegraphics[scale=0.03]{figs/map-logo-c.pdf},4*} \hspace{0.15cm} Ge Zhang \textsuperscript{\includegraphics[scale=0.03]{figs/map-logo-c.pdf},2,5*} \hspace{0.15cm} Xingran Chen $^6$  \hspace{0.15cm}  Hanzhi Yin$^3$ \hspace{0.15cm} Chenghua Lin$^{4\dag}$} 
{\bfseries{\small  
Emmanouil Benetos$^{1\dag}$ \hspace{0.15cm} Anton Ragni$^4$  \hspace{0.15cm} Norbert Gyenge$^4$ \hspace{0.15cm} Ruibo Liu$^7$\hspace{0.15cm} Gus Xia$^8$\hspace{0.15cm} Roger Dannenberg$^3$ \hspace{0.15cm} Yike Guo$^9$ \hspace{0.15cm} Jie Fu$^{2\dag}$ \hspace{0.15cm}} 
\\
\scriptsize
\textsuperscript{\includegraphics[scale=0.03]{figs/map-logo-c.pdf}} Multimodal Art Projection Research Community \hspace{0.15cm}
$^1$ Queen Mary University of London \hspace{0.15cm}
$^2$ Beijing Academy of Artificial Intelligence 
\\
\scriptsize
$^3$ Carnegie Mellon University \hspace{0.15cm}
$^4$ University of Sheffield
\hspace{0.2cm}
$^5$ University of Waterloo \hspace{0.15cm}
$^6$ University of Michigan Ann Arbor 
\\ \scriptsize
$^7$ Dartmouth College \hspace{0.15cm}
$^8$ New York University Shanghai\hspace{0.15cm}
$^9$ Hong Kong University of Science and Technology \\ \small
{\tt\scriptsize \{yinghao.ma, emmanouil.benetos\}@qmul.ac.uk, ruibiny@andrew.cmu.edu}
\\
{\tt\scriptsize \{yizhi.li, c.lin\}@sheffield.ac.uk, gezhang@umich.edu, fujie@baai.ac.cn}
}

\def\authorname{Y. Ma, R. Yuan, Y. Li, G. Zhang et al.}

\begin{document}

\maketitle

\begin{abstract}
Self-supervised learning (SSL) has shown promising results in various speech and natural language processing applications. However, its efficacy in music information retrieval (MIR) still remains largely unexplored. 
While previous SSL models pre-trained on music recordings may have been mostly closed-sourced, recent speech models such as wav2vec2.0 have shown promise in music modelling. 
Nevertheless, research exploring the effectiveness of applying speech SSL models to music recordings has been limited.
We explore the music adaption of SSL with two distinctive speech-related models, data2vec1.0 and Hubert, and refer to them as music2vec and musicHuBERT, respectively. 
We train $12$ SSL models with 95M parameters under various pre-training configurations and systematically evaluate the MIR task performances with 13 different MIR tasks. 
Our findings suggest that training with music data can generally improve performance on MIR tasks, even when models are trained using paradigms designed for speech.
However, we identify the limitations of such existing speech-oriented designs, especially in modelling polyphonic information.
Based on the experimental results, empirical suggestions are also given for designing future musical SSL strategies and paradigms.
\end{abstract}

\section{INTRODUCTION}

Deep learning (DL) techniques have shown promising results in a wide range of auditory tasks, including speech and music information retrieval (MIR). However, the quantity and quality of labelled data is a bottleneck for developing algorithms with better generalisation in complex real-world settings for machine listening. To address this issue, self-supervised learning (SSL) such as BERT \cite{kenton2019bert} has emerged as a solution to leverage diverse and representative unlabelled data to train a deep feature extractor with better generalisation. By combining this pre-trained SSL encoder with a naive classifier, typically a multi-layer perceptron (MLP) or long short-term memory (LSTM) with limited hidden layers, the model can achieve strong or state-of-the-art (SOTA) performance in various downstream tasks including NLP \cite{kenton2019bert, gururangan2020don, sarzynska2021detecting}, computer vision \cite{newell2020useful}, and audio \cite{baevski2020wav2vec, castellon2021codified}, where well-labelled datasets are limited. For music, larger datasets can be more expensive due to copyright and annotation costs, making SSL essential for developing effective MIR systems. Investigating versatile SSL approaches in MIR can further improve the performance on many MIR tasks, benefitting the music industry, music education, and heritage preservation.
Although SSL has significantly improved the performance of models in tasks such as speech recognition, sentiment analysis, and language modelling, its effectiveness in music information retrieval (MIR) remains largely unexplored.

There has been much work on SSL for audio representation learning, including speech, sound events or music. But most results are difficult to evaluate or fine-tune due to limited access to training data, pre-trained parameters or training codes. 
PANN \cite{kong2020panns} is trained on noisy/weak-label classification and does not provide promising results in music tasks such as pitch classification and instrument classification  \cite{ turian2022hear}. Besides, it can hardly be re-trained on music datasets given that the MIR community does not have a weekly labelled large music dataset.
MusiCoder \cite{zhao2021musicoder}, Music PASE \cite{wu2021multi}, and MAP-Music2Vec \cite{li2022map} use strategies mainly based on masked prediction, where training models predict the audio waveform manually-designed feature or learnable deep feature of input removed randomly from the ground truth. Such models trained on music are not open-sourced except MAP-Music2Vec, which provides pre-trained parameters on hugging-face\footnote{https://huggingface.co/m-a-p/music2vec-v1}. 
Jukebox \cite{dhariwal2020jukebox} uses similar strategies for pop-song recording generation and demonstrates good potential for multiple MIR tasks \cite{castellon2021codified}. But the training code for it is unavailable and is hard to fine-tune given its 6 billion parameters.
MAP-MERT v0 \cite{li2022large} mimicks HuBERT \cite{hsu2021hubert}, which regards the clustering results of audio as a pseudo label or pseudo spectrum to be reconstructed rather than a cluster assignment. But it does not provide training codes for further model evaluation.
Furthermore, there are some music SSL models based on instance discrimination. In this family of approaches, each instance is considered its class, and models are trained to distinguish among different instances. CLMR \cite{spijkervet2021contrastive} is trained with a limited number of parameters and shows limited capacity 
\cite{castellon2021codified}. PEMR \cite{yao2022contrastive} does not show promising results besides tagging and is not open-source for further evaluation. 

Although not designed for MIR tasks, some speech SSL models provide promising results on music tasks, and their training codes are available for fine-tuning or re-training on musical audio.
Mockingjay \cite{liu2020mockingjay}, and PASE \cite{ravanelli2020multi} use masked waveform / audio-feature prediction for pre-training.
COLR \cite{saeed2021contrastive} uses EfficientNet with a limited number of parameters and is designed for general audio, though it has a promising result on instrument classification. SF-NFNet-F0 \cite{wang2022towards} also uses an architecture based on convolution neural networks, a SlowFast Normalizer-Free ResNet, for audio pre-training.
Furthermore, apart from providing good results on automatic speech recognition (ASR), Wav2Vec2.0 \cite{baevski2020wav2vec}, HUBERT and data2vec \cite{baevski2022data2vec} also provide much better results on pitch estimation and instrumental classification than PANN, though they are still far from perfect \cite{turian2022hear}. 
All of the speech SSL models are helpful for music SSL model development.

Previous work on re-training speech SSL systems with music recordings is limited to the size of training datasets or model structure.
Ragano et al.~\cite{ragano2022learning} re-trained wav2vec2.0 on music audio and improved performance on pitch estimation and instrument classification significantly. But the training set is less than 100 hours which may be less representative, and the downstream tasks are limited and not universal. 
MusiCoder and Music PASE can be regarded as re-training speech SSL models on music recordings, but the model performance is not promising. Besides, these models are evaluated with a limited number of downstream tasks, making the learned embedding less persuasive. SF-NFNet-F0 is trained on music recordings and provides better results on multiple music tagging tasks \cite{mccallum2022supervised}. But its model architecture is based on CNN, without much room for further scale-up and longer sequence modelling. %

The missing science in the previous studies is as follows. All of the existing models trained on music are either with a limited number of parameters and capacity for MIR tasks other than tagging or not open-source for further evaluation. 
Some of the systems developed on speech or general audio recordings demonstrate promising but not satisfying results on MIR tasks. 
Besides, previous investigations on the efficacy of applying speech-related SSL models to music recordings are limited by the size of the training set, not enough universality on the downstream tasks, or paying less attention to powerful transformer structures.

Our key contributions are four-fold:
(1) exploring two speech-related SSL models based on transformer structures, data2vec and HuBERT, and comparing the results with those models pre-trained in speech recordings; (2) carrying out ablation studies for pre-training, thus providing more intuition for further music SSL system design; and (3) systematically comparing the performance on 13 downstream tasks, which facilitates comprehensive model evaluation on a wide range of MIR tasks. %


\begin{figure*}[!t]
  \centering

\begin{subfigure}{.45\textwidth}
  \centering
  \includegraphics[width=\linewidth]{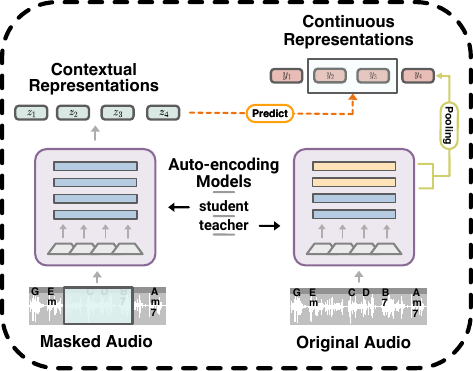}
  \caption{Music2Vec}
  \label{fig:framework_d2v}
\end{subfigure}
\hspace{5mm}
\begin{subfigure}{.45\textwidth}
  \centering
  \includegraphics[width=\linewidth]{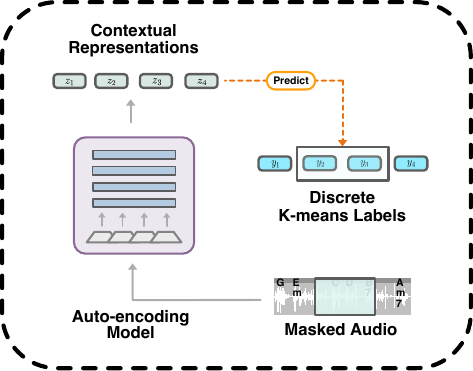}
  \caption{MusicHuBERT}
  \label{fig:framework_hb}
\end{subfigure}%
  \caption{
    Pre-training Paradigms of Selected Models. Both of the models are fed with masked audio inputs and predict given targets without supervised information.
  }\label{fig:framework}
\end{figure*}

\section{METHOD}
In order to keep the pre-training and representation evaluation protocols comparable, we focus on adapting from the speech self-supervised learning frameworks that support direct audio input and end-to-end pre-training.
Given our intent of exploring the influence of the pre-training design itself, we choose two SSL frameworks mainly distinguished by their self-supervised learning targets while sharing very similar training settings, including model architecture, training datasets, and evaluation protocols.
In this section, we briefly describe the two selected SSL models -- data2vec-1.0~\cite{baevski2022data2vec} and HuBERT~\cite{hsu2021hubert} -- in the unified auto-encoding framework (cf. Fig.~\ref{fig:framework}) and discuss the similarities and differences under music audio pre-training.

\subsection{Music2Vec: Continuous Target Prediction}
We adapt the pre-training paradigm from the speech version of the multi-modal framework data2vec-1.0~\cite{baevski2022data2vec}, where the prediction targets during pre-training are continuous representations. We refer to this continuous prediction model adapted with music recordings as Music2Vec.

Modified from the design of bootstrap your own latent (BYOL)~\cite{grill2020bootstrap}, Music2Vec aims to predict continuous latent representations from the teacher model for the masked input audios, which is illustrated in Fig.~\ref{fig:framework_d2v}. 
The teacher model and student model share the same architecture, and the parameters of the teacher model are updated according to the exponential moving average of the student \cite{baevski2022data2vec}. 
The student model takes the partially masked input and is asked to predict the average pooling of top-$K$ layer outputs from the Transformer~\cite{vaswani2017attention} in the teacher model. 
In contrast, the teacher model takes the unmasked input and provides contextual prediction targets in the pre-training.

Following the data2vec~\cite{baevski2020wav2vec} setting, we train the Music2Vec of 95M parameters with a comparable 1k hours of music recordings. 
Since pre-trained speech models can barely benefit music representation learning \cite{ragano2022learning}, we instead train the base model from scratch to verify its effectiveness in modelling music audio recordings.

\subsection{MusicHuBERT: Discrete Target Prediction}
Another efficient speech SSL model, HuBERT~\cite{hsu2021hubert}, is chosen as the representative of discrete target prediction design. We referred to the music adaption version as MusicHuBERT.

The MusicHuBERT model takes masked music audios as input (Similar to Music2Vec) and predicts pre-processed discrete labels corresponding to the masked area, as shown in Fig.~\ref{fig:framework_hb}. The discrete targets are pseudo labels provided by K-means that are trained on the MFCC features of the training audios. The number of clusters $K$ of the K-means model is a hyperparameter, and all the centroids are assigned with randomly initialised embeddings and learned during the MusicHuBERT pre-training. MusicHuBERT can also be trained for an extra $n$ iterations, where K-means clustering is learned from model outputs' previous iteration. We follow the original HuBERT~\cite{hsu2021hubert} setting to train a model with 95M parameters of the same size as Music2Vec.

\subsection{similarities \& Differences of SSL frameworks} 
This subsection will examine the similarities and differences between the SSL frameworks mentioned above.

Both Music2Vec and MusicHuBERT are annotation-free and utilise SSL techniques; their most common characteristic is the training task of ``reconstructing'' information from masked inputs, making them auto-encoding models. 
During the denoising process, these models learn the semantics contained in the audio. Furthermore, they share similar model architecture designs, which are inherited from wav2vec-2.0~\cite{baevski2020wav2vec}, wherein the audio is initially encoded by a multi-layer 1-D CNN feature extractor that maps a 16 kHz waveform to 50 Hz representations. The encoded tokens are then fed into a 12-layer transformer block with a hidden dimension of $H=768$. %

Regarding the differences in the designs, the most notable one is that Music2Vec is required to predict continuous latent
variables, whereas MusicHuBERT predicts discrete pseudo-labels. The time cost of SSL target preparation bottleneck varies according to their mechanism. In Music2Vec, the pre-training consumes twice the model forward time since the target representations from the teacher model are inferred on-the-fly. In contrast, MusicHuBERT trains the K-means model and infers all the pseudo-labels before training, which requires high parallel processing ability when the dataset is scaled-up.

\section{DATASET \& EVALUATION}
\subsection{Training}
We use a private dataset with 1000 hours of music audio recordings for pre-training; each sample is a 30s-long excerpt from pop-song or instrumental music. The size of the pre-training dataset is roughly the same as the pre-training for HuBERT-base and data2vec-audio-base models.

\subsection{Evaluation}
We evaluate the models on 13 downstream tasks, including timbre classification tasks such as genre and instrumental classification, singing, playing technique classification, singer classification, and music tagging; emotion-related tasks like music mood classification and regression; and note-related tasks such as pitch estimation, key detection; and sequential tasks like beat tracking.

\textbf{Music Tagging} is a multi-label classification task. We used MagnaTagATune (MTT) \cite{law2009evaluation} and MTG-Jamendo \cite{bogdanov2019mtg} for this task, tag categories of which include genre, instrumentation, mood, and tempo (e.g. fast) etc. For both datasets, we limit the tag vocabulary to the 50 most common tags. We use all clips in MTT and MTG-Jamendo for evaluation. Since many of the audio recordings among 5.5k MTG-Jamendo excerpts are longer than the 30s, we averaged the multiple embeddings computed with a 30s sliding window as the overall embedding. The metrics are the macro-average of ROC-AUCs and the average precision (AP) / PR-AUC among all top-50 tags.

\textbf{Key detection}. We use a commonly-used subset of Giantsteps-MTG-keys \cite{korzeniowski2017end} as
the training and validation set following the data splitting \cite{castellon2021codified}, and Giantsteps (GS) \cite{knees2015two} as the test set. The metric is a refined accuracy that gives partial credit to reasonable errors \cite{raffel2014mir_eval}.

\textbf{Genre classification}. We report the multi-class classification accuracy of the GTZAN \cite{tzanetakis2002musical} dataset, along with ROC and AP on MTG-Genre for multi-label. We used the standard "fail-filtered" split \cite{kereliuk2015deep} for GTZAN.

\textbf{Emotion score regression.} The Emomusic dataset \cite{soleymani20131000} contains 744 music clips of 45 seconds, each reported on a two-dimensional valence-arousal plane after listening. %
We use the same dataset split as \cite{castellon2021codified}. The official evaluation metric is the determination coefficient ($r^2$) between the model regression results and human annotations of arousal (EmoA) and valence (EmoV) \cite{soleymani20131000}. We split the 45-second clip into a 5-second sliding window for inference and averaged the prediction. 

\textbf{Instrument classification}. We use the Nsynth \cite{engel2017neural} and MTG-instrument datasets. The former is a multi-class task on 306k audio samples in 11 instruments with accuracy as an indicator. The latter is a subset of MTG-Jamendo, containing 25k audio tracks and 41 instrument tags; each track can contain multiple instruments and is evaluated on ROC and AP.

\textbf{Pitch classification}. Given these audios are short monophonic audio, this task is multi-class to determine which of the 128 pitch categories, and the accuracy is used as an evaluation metric. 

\textbf{Vocal technique detection}. We use the VocalSet dataset \cite{wilkins2018vocalset}, which is the only publicly available dataset for the study of singing techniques. The dataset contains the vocals of 17 different singing techniques in various contexts for a total of 10.1 hours. As the audio clips are divided into 3 seconds, the task only requires a judgment on the type of technique and not on the start and end of the technique. We used the same 10 different singing techniques as in \cite{yamamoto2022deformable} as a subset and used the same 15 singers as the training and validation sets and 5 singers as the test set. Since there is no accepted division between training and validation sets, we selected 9 singers as the training set and 6 singers as the validation set. All the 3-second segments originate from the same recording are allocated to the same part of the split (e.g. all in the testing set). 

\textbf{Singer identification} is to identify the vocal performer from a given recording. We randomly divided the VocalSet dataset, which contains 20 different professional singers (9 female and 11 male), into a training set, validation set and testing set based on a ratio of 12:8:5, all containing the same 20 singers.

\textbf{Beat tracking}. We use an offline approach to the binary classification, i.e. the model can use the following information from each frame to help with inference. The model needs to output frame-by-frame predictions at a certain frequency and post-process them using a dynamic Bayesian network (DBN) \cite{bock2016joint}, the same methods with supervised SOTA. The DBN is implemented using \texttt{madmom} \cite{madmom}. The dataset we use is GTZAN Rhythm \cite{marchand2015swing}. We also label the two adjacent frames of each label as beat, a common way of smoothing in beat tracking. The model is evaluated using the f\_measure implemented in \texttt{mir\_eval} \cite{raffel2014mir_eval}, and the prediction is considered correct if the difference between the predicted event and the ground truth does not exceed 20ms. In this task, some models were trained on other datasets, and the full GTZAN set was used as the test set. For all cases, however, we use GTZAN-train as the training set and GTZAN-test as the test set.

\textbf{Emotion Tagging.} We use MTG-MoodTheme, another subset of MTG-Jamendo \cite{bogdanov2019mtg} that contains 18.5k audio tracks and 59 tags. Unlike Emomusic, this is a multi-label task, with ROC and AP as metrics.

\section{EXPERIMENTAL RESULTS}
We use the \texttt{fairseq} framework\footnote{https://github.com/facebookresearch/fairseq} from Meta to train MusicHuBERT and Music2Vec models. All the MusicHuBERT and Music2Vec models are trained for 400k steps with 8 $\times$ NVIDIA A100-40GB GPUs. 
Training with 8 GPUs takes around $2-3$ days. 
The experimental results are chiefly as follows.

Our findings suggest these SSL models pre-trained on speech can be helpful for MIR tasks, but pre-trained on music is generally more helpful, besides some exceptions. In section 4.2, we identify the strengths along with weaknesses of training strategies, revealing areas for further improvement. In section 4.3, we discuss the effect of hyperparameters in pretext tasks.

\subsection{Pre-trained on Speech and Music}
\begin{table*}[htb]
    \centering
    \setlength{\tabcolsep}{3pt}
    \caption{Experimental performance of the SSL baseline systems on all downstream tasks}\label{table:results}
    \resizebox{\textwidth}{!}{
    \begin{tabular}{l|cccccccccccccccccccc}
    \hline
    Downstream & \multicolumn{2}{c}{MTT}&\multirow{2}{*}{GS key} &GTZAN & \multicolumn{2}{c}{EMO}  &Nsynth & Nsynth & VocalSet & VocalSet & GTzAN &\multicolumn{2}{c}{MTG} &\multicolumn{2}{c}{MTG} &\multicolumn{2}{c}{MTG}  & \multicolumn{2}{c}{MTG} \\ 
    dataset & & & &Genre & & &Instr &pitch &tech &singer &Rhythm &\multicolumn{2}{c}{Instrument} &\multicolumn{2}{c}{MoodTheme} &\multicolumn{2}{c}{Genre} &\multicolumn{2}{c}{Top50} \\ \hline
    Metrics &ROC &AP &Refined Acc &Acc & $Emo_V$ & $Emo_A$ &Acc &Acc &Acc &Acc & F1 (beat) &ROC &AP  &ROC &AP  &ROC &AP &ROC &AP \\ \hline \hline
    HuBERT &\multirow{2}{*}{89.8} &\multirow{2}{*}{36.4} &\multirow{2}{*}{15.0} &\multirow{2}{*}{64.8} &\multirow{2}{*}{31.0} &\multirow{2}{*}{57.5} &\multirow{2}{*}{68.2} &\multirow{2}{*}{79.4} &\multirow{2}{*}{61.0} &\multirow{2}{*}{58.8} & \multirow{2}{*}{83.5} &\multirow{2}{*}{73.2} &\multirow{2}{*}{17.0}  &\multirow{2}{*}{74.0} &\multirow{2}{*}{11.6}  &\multirow{2}{*}{85.0} &\multirow{2}{*}{16.3} &\multirow{2}{*}{81.8} &\multirow{2}{*}{26.5} \\
   base& & & & & & & & & & & & & & & & & & & \\ 
   \hline
    MusicHuBERT &\multirow{2}{*}{\underline{90.2}} &\multirow{2}{*}{\underline{37.7}} &\multirow{2}{*}{14.7} &\multirow{2}{*}{\underline{70.0}} &\multirow{2}{*}{\underline{42.1}} &\multirow{2}{*}{\underline{66.5}} &\multirow{2}{*}{69.3} &\multirow{2}{*}{77.4} &\multirow{2}{*}{65.9} &\multirow{2}{*}{\underline{75.3}} & \multirow{2}{*}{\underline{\textbf{88.6}}} &\multirow{2}{*}{\underline{75.5}} &\multirow{2}{*}{\underline{17.8}}  &\multirow{2}{*}{\underline{76.0}} &\multirow{2}{*}{\underline{13.9}}  &\multirow{2}{*}{\underline{86.5}} &\multirow{2}{*}{\underline{18.0}} &\multirow{2}{*}{\underline{82.4}} &\multirow{2}{*}{\underline{28.1}} \\
   base& & & & & & & & & & & & & & & & & & & \\ 
   \hline\hline
    data2vec &\multirow{2}{*}{88.4} &\multirow{2}{*}{33.6} &\multirow{2}{*}{15.5} &\multirow{2}{*}{60.7} &\multirow{2}{*}{23.0} &\multirow{2}{*}{49.6} &\multirow{2}{*}{69.3} &\multirow{2}{*}{77.7} &\multirow{2}{*}{64.9} &\multirow{2}{*}{74.6} & \multirow{2}{*}{36.4} &\multirow{2}{*}{73.1} &\multirow{2}{*}{16.9}  &\multirow{2}{*}{73.3} &\multirow{2}{*}{11.0}  &\multirow{2}{*}{83.5} &\multirow{2}{*}{14.5} &\multirow{2}{*}{80.6} &\multirow{2}{*}{24.8}  \\
   audio base& & & & & & & & & & & & & & & & & & & \\ 
   \hline
    Music2vec &\multirow{2}{*}{89.1} &\multirow{2}{*}{35.1} &\multirow{2}{*}{\underline{19.0}} &\multirow{2}{*}{59.7} &\multirow{2}{*}{38.5} &\multirow{2}{*}{61.9} &\multirow{2}{*}{\underline{69.4}} &\multirow{2}{*}{\underline{88.9}} &\multirow{2}{*}{\textbf{\underline{68.3}}} &\multirow{2}{*}{69.5} & \multirow{2}{*}{33.5} &\multirow{2}{*}{73.1} &\multirow{2}{*}{16.3}  &\multirow{2}{*}{74.3} &\multirow{2}{*}{12.2}  &\multirow{2}{*}{85.2} &\multirow{2}{*}{16.5} &\multirow{2}{*}{81.4} &\multirow{2}{*}{26.2}  \\
   vanilla& & & & & & & & & & & & & & & & & & & \\ \hline
   \hline
   \multirow{2}{*}{SOTA} &\multirow{2}{*}{\textbf{92.0}\cite{huang2020large}} &\multirow{2}{*}{\textbf{41.4}\cite{castellon2021codified}} &\multirow{2}{*}{\textbf{74.3}\cite{korzeniowski2017end}} &\multirow{2}{*}{\textbf{82.1}\cite{lee2018samplecnn}} &\multirow{2}{*}{\textbf{61.7}} &\multirow{2}{*}{\textbf{72.1}\cite{castellon2021codified}} &\multirow{2}{*}{\textbf{78.2}\cite{wang2022towards}} &\multirow{2}{*}{\textbf{89.2}\cite{mccallum2022supervised}} &\multirow{2}{*}{65.6\cite{yamamoto2022deformable}} &\multirow{2}{*}{\textbf{80.3}\cite{modrzejewski2023transfer}} & \multirow{2}{*}{80.6\cite{heydari2021beatnet}} &\multirow{2}{*}{\textbf{78.8}} &\multirow{2}{*}{\textbf{20.2}\cite{alonso2022music}}  &\multirow{2}{*}{\textbf{78.6}} &\multirow{2}{*}{\textbf{16.1}\cite{mccallum2022supervised}}  &\multirow{2}{*}{\textbf{87.7}} &\multirow{2}{*}{\textbf{20.3}\cite{alonso2022music}} &\multirow{2}{*}{\textbf{84.3}} &\multirow{2}{*}{\textbf{32.1}\cite{mccallum2022supervised}}  \\
   & & & & & & & & & & & & & & & & & & & \\ 
   \hline
    \end{tabular}
    }
\end{table*}

Table \ref{table:results} demonstrates the performance of HuBERT\footnote{https://huggingface.co/facebook/HuBERT-base-ls960} and data2vec\footnote{https://huggingface.co/facebook/data2vec-audio-base} SSL models that were pre-trained on speech recordings and music recordings separately. Here, we only consider the SOTA performance trained with the same dataset train/valid/test split.
All of the models are used as parameter-frozen feature extractors. The weighted sum of one output of the CNN tokeniser as well as the 12 outputs of all the transformer layers, are combined with an MLP as the back end. The MLP has only one single 512-dimension hidden layer. The learning rate of the probing is set to 1e-3. 

For the HuBERT model, the results pre-trained on speech recordings are comparable with SOTA on tasks like music tagging, beat tracking, pitch estimation and singing technique classification etc., and are surpassed by the results pre-trained on music audio on most of the downstream tasks besides pitch estimation on Nsynth and key detection on GS. 
For pitch detection, the data samples in Nsynth are a single note played by one single monophonic instrument, which is similar to speech data. So it is reasonable that HuBERT pre-trained on speech data is capable of modelling a single pitch. Although HuBERT surpasses the vanilla MusicHuBERT on GS and Nsynth-pitch, it is surpassed by the results of MusicHuBERT with an ablation study on pre-training hyperparameters (shown in Table \ref{table:muhubert}). 

For data2vec, the data2vec-audio results are also comparable with SOTA on many tasks and have a large gap on others, and overall surpassed by Music2Vec or its ablation study shown in Table \ref{table:music2vec} on most of the tasks as well. But the data2vec results of beat tracking on GTZAN-Rhythm and singer identification on Vocalset surpassed all Music2Vec. Vocalset includes singing of different phonemes with different singing techniques by different singers. The speech SSL system is capable of modelling diverse phonemes in ASR and various timbres of speakers but has less focus on timbre in speaking techniques you may find in opera. On the contrary, the music SSL models may focus more on phonemes (lyrics) and singing timbre (techniques) but include less focus on the singer itself. For beat tracking, we observe that the performance is reduced significantly when the number of transformer layers increases from 0 to 12. This shows that the data2vec structure may not be useful for learning temporal information.

\subsection{Pre-trained with Different Paradigms}
From Table \ref{table:results}, we can tell that MusicHuBERT is more promising than Music2vec  given that it provides better results in most of the downstream tasks, especially genre classification on GTZAN, emotion regression on EMO and beat tracking on GTZAN. But it is worse on single-pitch estimation on Nsynth, along with key detection on GS.

These phenomena suggest pre-training with the HuBERT paradigm is strongly correlated with the MFCC feature information used for k-means. Therefore, the quantisation results lack multi-pitch information, including harmony or chord modelling, that is essential to key detection. The following research can use the chroma feature to replace MFCCs\footnote{For more information on this, please refer to our following paper MERT: Acoustic Music Understanding Model with Large-Scale Self-supervised Training at \url{https://arxiv.org/abs/2306.00107}}.
On the contrary, the mask prediction for the deep feature in the data2vec pre-training paradigm is clearly better but still has much room for improvement compared to the SOTA. Although the deep feature still lacks sufficient harmonic information for key detection, it already contains enough information for single-pitch estimation, and the MFCCs may focus more on the timbre of instruments instead of the fundamental frequency. Apparently, Music2Vec can learn pitch information more freely. 
Besides, data2vec is generally a bit worse for tagging than Music2vec, and both are significantly worse on beat tracking compared to HuBERT and MusicHuBERT.

\subsection{Ablation Studies on Pretraining Hyperparameters}
Here, we carry out an ablation study of hyperparameter search under both pre-training paradigms. Given the time limitation, we did not extract features on MTG datasets and only calculated the results in another 9 downstream tasks. 
\subsubsection{Ablation Study on MusicHuBERT}

\begin{table*}[htb]
    \centering
    \setlength{\tabcolsep}{3pt}

    \caption{Ablation study on MusicHuBERT hyperparameters (k is the number of MFCC clusters)}
    \resizebox{0.94\textwidth}{!}{
    \begin{tabular}{|l|ccccccccccc|c|}
    \hline
    Downstream & \multicolumn{2}{c}{MTT}&\multirow{2}{*}{GS key} &GTZAN & \multicolumn{2}{c}{EMO}  &Nsynth &Nsynth & VocalSet & VocalSet &GTZAN & Average  \\ 
    dataset & & & &Genre & & &Instr &pitch &tech &singer &Rhythm & Score \\ \hline
    Metrics &ROC &AP &Refined Acc &Acc & $Emo_V$ & $Emo_A$ &Acc &Acc &Acc &Acc &F1 (beat) & score\\ \hline \hline

    HuBERT &89.8 &36.4 &15.0 &64.8 &31.0 &57.5 &68.2 &79.4 &61.0&58.8 &83.5 & 59.8\\    \hline

    k=2000 MFCC dim=39 &90.2 &37.7 &14.7 &\textbf{70.0} &42.1 &66.5 &69.3 &77.4 &65.9 &75.3 &88.6 &64.4\\ 
    k=2000 iter2 &\textbf{90.4} &37.5 &13.8 &68.3 &\textbf{43.3} &67.4 &70.0 &\textbf{80.3} &63.6 &70.4 &\textbf{88.8} &63.8\\   
    k=500 MFCC dim=39 &89.6 &36.1 &15.7 &64.5 &41.0 &67.7 &66.7 &76.8 &60.5 &72.3 &87.5 &62.4\\ 
    k=500 MFCC dim=60 &90.3 &\textbf{38.0} &\textbf{17.6} &69.7 &40.8 &\textbf{67.5} &\textbf{70.3} &79.0 &\textbf{66.2} &\textbf{75.5} &88.6 &\textbf{65.0}\\ 
    \hline
    \end{tabular}
    }
    \label{table:muhubert}
\end{table*}

 We use the number of clusters k =500 and k=2000. 
For the case k=500, we increase the dimension of MFCC features from 13, which is commonly used in the speech community, to 20, which is widely used in sound event detection. Thus, the dimension of MFCCs combined with their delta features and delta-delta features have 39 and 60 dimensions respectively. 
For the case of k=2000, we use the 768-dimension deep feature learned from the first iteration experiment to carry out the second iteration k-means. 

From Table \ref{table:muhubert}, we can see that MusicHuBERT with k=2000 is better than the k=500 case for most of the tasks. Given HuBERT is good for speech when k=100 or k=500, which is roughly the number of human phonemes, this implies music tokens or notes are much richer than speech and therefore need a larger number for quantisation.

The results on k-means for deep features are better than the vanilla MusicHuBERT besides genre classification on GTZAN, singer identification on vocalset, and singing techniques classification on vocalset. This implies the MFCCs features are good for modelling the human voice, regardless of speech or singing. The results of GTZAN may be due to the randomness as the dataset is  very small.

Besides, increasing the dimension of MFCCs provides no significant difference among most of the tasks other than tasks on Nsynth and GS. Increased dimensionality for MFCC features can provide more detailed information on impulse response for a sound event. Thus,  monophonic instrumental notes can be better modelled with 60-dimension MFCC features. Furthermore, the emotion regression also provides different results, but the average of the two metrics is nearly the same, providing no significant improvement.

\subsubsection{Ablation Study on Music2Vec}

\begin{table*}[htb]
    \centering
    \setlength{\tabcolsep}{3pt}
    \caption{Ablation study on Music2Vec hyperparameters (span is mask span, prob is mask probability, step is training steps, target=12 uses all 12 transformer layers, and crop5s uses 5s music excerpts)}
    \resizebox{0.9\textwidth}{!}{
    \begin{tabular}{|l|ccccccccccc|c|}
    \hline
    Downstream & \multicolumn{2}{c}{MTT}&\multirow{2}{*}{GS key} &GTZAN & \multicolumn{2}{c}{EMO}  &Nsynth &Nsynth & VocalSet & VocalSet &GTZAN &Average \\ 
    dataset & & & &Genre & & &Instr &pitch &tech &singer &Rhythm &Score\\ \hline
    Metrics &ROC &AP &Refined Acc &Acc & $Emo_V$ & $Emo_A$ &Acc &Acc &Acc &Acc &F1 (beat) & score\\ \hline 
    \hline
    data2vec &88.4 &33.6 &15.5 &60.7 &23.0 &49.6 &69.3 &77.7 &64.9 &\textbf{74.6} &\textbf{36.4} & 55.2\\  \hline 

     vanilla&89.1 &35.1 &19.0 &59.7 &38.5 &61.9 &69.4 &88.9 &68.3 &69.5 &33.5 &57.8\\ 
     span=5  &87.3 &32.0 &15.7 &47.6 &22.7 &41.2 &64.2 &84.8 &56.7 &53.8 &33.2 &49.7\\   
     span=15  &88.7 &34.3 &16.4 &56.6 &39.0 &58.8 &67.1 &88.1 &63.1 &61.9 &33.1 &55.2\\ 
     prob=50 &88.5 &34.0 &23.7 &59.3 &40.6 &55.0 &66.8 &87.7 &64.9 &61.7 &33.9 &56.3\\ 
     prob=80 &88.2 &33.9 &18.4 &50.3 &36.7 &55.7 &67.9 &88.9 &64.2 &65.2 &33.7 &55.1\\ 
     step=800k &87.7 &32.7 &20.3 &54.5 &34.9 &47.3 &66.9 &87.5 &65.6 &65.1 &33.4 &55.0\\   
     target=12 &89.7 &35.2 &\textbf{26.5} &64.5 &41.7 &64.2 &\textbf{71.1} &\textbf{89.2} &71.0 &73.2 &34.1 &60.6\\ 
     crop5s&\textbf{90.0} &\textbf{36.6} &18.5 &\textbf{76.6} &\textbf{53.4} &\textbf{71.6} &68.3 &88.9 &\textbf{71.3} &72.4 &33.9 &\textbf{61.8}\\ 
    \hline
    \end{tabular}
}
    \label{table:music2vec}
\end{table*}

We use audio files with 30s length, mask span length 10, mask probability 65\%, target top-$8$ transformer layer the teacher model as a deep feature, and training step 400K as the vanilla setting. We conduct parameter searching and correlation analysis for Music2Vec pretraining, including masking strategy, training steps, the learning target layers, and recording length; the results are shown in Table \ref{table:music2vec}.

We revise the masking strategy by changing the \textbf{mask span length} and \textbf{mask token probability} in the data2vec-audio-base setting. 
Mask token probability is the probability for each token to be chosen as the start of the span to be masked, the length of which can also be adapted for different data modalities. The results in Table \ref{table:music2vec} show that the other span value and other mask token probability provide a bit worse results on nearly all the tasks. This suggests that the data2vec hyperparameters for speech pre-training are generally helpful for music pre-training. 

Given the fact that early transformer layer representations generally perform well on key detection and beat tracking, we modify the \textbf{prediction target} provided by the teacher model. We change the prediction target in Music2Vec from the original one, that is, the average of the top-8 layer representations, to all the 12 layers. The results in Table \ref{table:music2vec} show that Music2Vec actually benefits, not only from the potentially preserved key information shown by a significant increase on GS but all the other tasks as well. 

Furthermore, we use \textbf{audio length cropping} to shorten music excerpts since longer sequences are more difficult to model.
Note that the combined audio length in a batch on a single GPU is not altered, and the hardware environment remains the same, making a single training batch contain a larger number of music samples when clips are cropped. Due to the position embedding in the SSL systems, the model can get information more than 5 seconds after pre-training on only 5-second music recordings. But the key detection provides worse results which may lead to the fact that a local key within a 5-second song may not be identical to the global key in the whole music sentence.

\section{CONCLUSION \& DISCUSSION}
In this paper, we explore the music variants of two distinctive speech-related transformer-based SSL models, data2vec and HuBERT. Our findings suggest that pre-training with music recordings rather than speech can generally improve performance on a wide range of MIR tasks, even when the models and training are designed for speech. There are exceptions for data2vec, however, such as singer identification, the dataset of which is similar to the speech dataset used to pre-train. Thus, when resources are limited, our suggestion is to use speech pre-training models, given that they can provide helpful information about music already. Speech data can be beneficial if lacking a sufficient vocal dataset with different singers, but one should use mainly music data if possible. 

Furthermore, we can use the same speech training hyperparameters for masked span and masked probability in music pre-training. But some other hyperparameters, such as the number for pseudo label clustering, might be the shortage of pretext strategies. We identified some limitations of existing speech SSL systems, especially in the case of harmonic information and diversity of music notes. One suggestion is to emphasise key or harmonic in the pretext task for music SSL models by using more than just MFCC features. Also, the number of categories for quantisation in k-means should be much larger if necessary, given the number of pitch, chord, and timbre categories is much larger than the number of human speech phones. This diversity in music might be a bottleneck for both speech SSL systems to learn good music features. For one thing, the larger number of clusters for k-means in HuBERT is expensive to calculate, making it harder to scale up, preventing transformer-based models from reaching their potential for better performance and longer sequence modelling. In addition, it may not be easy for data2vec to jointly learn deeper features. We may need curriculum learning skills or manually-designed features to increase training stability. 

Another general suggestion for pre-training recognises that batch size should be as diverse as possible. Given that the memory of one single machine is limited, it is a good idea to shorten the length of audio to be modelled at first, allowing for an increase in batch size, and then train another language model for long sequence modelling.

We believe the findings in this paper to be of value in understanding the potential for SSL speech models applied to music, and we have offered some general insights about music modelling that resulted from this study.

\section{Acknowledgements}
Yinghao Ma is a research student at the UKRI Centre for Doctoral Training in Artificial Intelligence and Music, supported by UK Research and Innovation [grant number EP/S022694/1]. 
Yizhi Li is fully funded by an industrial PhD studentship (Grant number: 171362) from the University of Sheffield, UK. 
This work is supported by the National Key R\&D Program of China (2020AAA0105200). 
We acknowledge IT Services at The University of Sheffield for the provision of services for High-Performance Computing.
We would also like to express great appreciation for the suggestions from faculties Dr Chris Donahue, and Dr Roger Dannenberg, as well as the facility support from Mr. Yulong Zhang in the preliminary stage.

\bibliography{reference}

\begin{thebibliography}{10}
\providecommand{\url}[1]{#1}
\csname url@samestyle\endcsname
\providecommand{\newblock}{\relax}
\providecommand{\bibinfo}[2]{#2}
\providecommand{\BIBentrySTDinterwordspacing}{\spaceskip=0pt\relax}
\providecommand{\BIBentryALTinterwordstretchfactor}{4}
\providecommand{\BIBentryALTinterwordspacing}{\spaceskip=\fontdimen2\font plus
\BIBentryALTinterwordstretchfactor\fontdimen3\font minus
  \fontdimen4\font\relax}
\providecommand{\BIBforeignlanguage}[2]{{%
\expandafter\ifx\csname l@#1\endcsname\relax
\typeout{** WARNING: IEEEtran.bst: No hyphenation pattern has been}%
\typeout{** loaded for the language `#1'. Using the pattern for}%
\typeout{** the default language instead.}%
\else
\language=\csname l@#1\endcsname
\fi
#2}}
\providecommand{\BIBdecl}{\relax}
\BIBdecl

\bibitem{kenton2019bert}
J.~D. M.-W.~C. Kenton and L.~K. Toutanova, ``Bert: Pre-training of deep
  bidirectional transformers for language understanding,'' in \emph{Proceedings
  of NAACL-HLT}, 2019, pp. 4171--4186.

\bibitem{gururangan2020don}
S.~Gururangan, A.~Marasovi{\'c}, S.~Swayamdipta, K.~Lo, I.~Beltagy, D.~Downey,
  and N.~A. Smith, ``Don't stop pretraining: Adapt language models to domains
  and tasks,'' \emph{arXiv preprint arXiv:2004.10964}, 2020.

\bibitem{sarzynska2021detecting}
J.~Sarzynska-Wawer, A.~Wawer, A.~Pawlak, J.~Szymanowska, I.~Stefaniak,
  M.~Jarkiewicz, and L.~Okruszek, ``Detecting formal thought disorder by deep
  contextualized word representations,'' \emph{Psychiatry Research}, vol. 304,
  p. 114135, 2021.

\bibitem{newell2020useful}
A.~Newell and J.~Deng, ``How useful is self-supervised pretraining for visual
  tasks?'' in \emph{Proceedings of the IEEE/CVF Conference on Computer Vision
  and Pattern Recognition}, 2020, pp. 7345--7354.

\bibitem{baevski2020wav2vec}
A.~Baevski, Y.~Zhou, A.~Mohamed, and M.~Auli, ``wav2vec 2.0: A framework for
  self-supervised learning of speech representations,'' \emph{Advances in
  neural information processing systems}, vol.~33, pp. 12\,449--12\,460, 2020.

\bibitem{castellon2021codified}
R.~Castellon, C.~Donahue, and P.~Liang, ``Codified audio language modeling
  learns useful representations for music information retrieval,'' \emph{arXiv
  preprint arXiv:2107.05677}, 2021.

\bibitem{kong2020panns}
Q.~Kong, Y.~Cao, T.~Iqbal, Y.~Wang, W.~Wang, and M.~D. Plumbley, ``Panns:
  Large-scale pretrained audio neural networks for audio pattern recognition,''
  \emph{IEEE/ACM Transactions on Audio, Speech, and Language Processing},
  vol.~28, pp. 2880--2894, 2020.

\bibitem{turian2022hear}
J.~Turian, J.~Shier, H.~R. Khan, B.~Raj, B.~W. Schuller, C.~J. Steinmetz,
  C.~Malloy, G.~Tzanetakis, G.~Velarde, K.~McNally \emph{et~al.}, ``Hear:
  Holistic evaluation of audio representations,'' in \emph{NeurIPS 2021
  Competitions and Demonstrations Track}.\hskip 1em plus 0.5em minus
  0.4em\relax PMLR, 2022, pp. 125--145.

\bibitem{zhao2021musicoder}
Y.~Zhao and J.~Guo, ``Musicoder: A universal music-acoustic encoder based on
  transformer,'' in \emph{International Conference on Multimedia
  Modeling}.\hskip 1em plus 0.5em minus 0.4em\relax Springer, 2021, pp.
  417--429.

\bibitem{wu2021multi}
H.-H. Wu, C.-C. Kao, Q.~Tang, M.~Sun, B.~McFee, J.~P. Bello, and C.~Wang,
  ``Multi-task self-supervised pre-training for music classification,'' in
  \emph{ICASSP 2021-2021 IEEE International Conference on Acoustics, Speech and
  Signal Processing (ICASSP)}.\hskip 1em plus 0.5em minus 0.4em\relax IEEE,
  2021, pp. 556--560.

\bibitem{li2022map}
Y.~Li, R.~Yuan, G.~Zhang, Y.~MA, C.~Lin, X.~Chen, A.~Ragni, H.~Yin, Z.~Hu,
  H.~He \emph{et~al.}, ``Map-music2vec: A simple and effective baseline for
  self-supervised music audio representation learning,'' in \emph{ISMIR late
  braking demo}, 2022.

\bibitem{dhariwal2020jukebox}
P.~Dhariwal, H.~Jun, C.~Payne, J.~W. Kim, A.~Radford, and I.~Sutskever,
  ``Jukebox: A generative model for music,'' \emph{arXiv preprint
  arXiv:2005.00341}, 2020.

\bibitem{li2022large}
Y.~Li, R.~Yuan, G.~Zhang, Y.~Ma, C.~Lin, X.~Chen, A.~Ragni, H.~Yin, Z.~Hu,
  H.~He \emph{et~al.}, ``Large-scale pretrained model for self-supervised music
  audio representation learning,'' in \emph{Digital Music Research Network
  (DMRN) workshop}, 2022.

\bibitem{hsu2021hubert}
W.-N. Hsu, B.~Bolte, Y.-H.~H. Tsai, K.~Lakhotia, R.~Salakhutdinov, and
  A.~Mohamed, ``Hubert: Self-supervised speech representation learning by
  masked prediction of hidden units,'' \emph{IEEE/ACM Transactions on Audio,
  Speech, and Language Processing}, vol.~29, pp. 3451--3460, 2021.

\bibitem{spijkervet2021contrastive}
J.~Spijkervet and J.~A. Burgoyne, ``Contrastive learning of musical
  representations,'' \emph{arXiv preprint arXiv:2103.09410}, 2021.

\bibitem{yao2022contrastive}
D.~Yao, Z.~Zhao, S.~Zhang, J.~Zhu, Y.~Zhu, R.~Zhang, and X.~He, ``Contrastive
  learning with positive-negative frame mask for music representation,'' in
  \emph{Proceedings of the ACM Web Conference 2022}, 2022, pp. 2906--2915.

\bibitem{liu2020mockingjay}
A.~T. Liu, S.-w. Yang, P.-H. Chi, P.-c. Hsu, and H.-y. Lee, ``Mockingjay:
  Unsupervised speech representation learning with deep bidirectional
  transformer encoders,'' in \emph{ICASSP 2020-2020 IEEE International
  Conference on Acoustics, Speech and Signal Processing (ICASSP)}.\hskip 1em
  plus 0.5em minus 0.4em\relax IEEE, 2020, pp. 6419--6423.

\bibitem{ravanelli2020multi}
M.~Ravanelli, J.~Zhong, S.~Pascual, P.~Swietojanski, J.~Monteiro, J.~Trmal, and
  Y.~Bengio, ``Multi-task self-supervised learning for robust speech
  recognition,'' in \emph{ICASSP 2020-2020 IEEE International Conference on
  Acoustics, Speech and Signal Processing (ICASSP)}.\hskip 1em plus 0.5em minus
  0.4em\relax IEEE, 2020, pp. 6989--6993.

\bibitem{saeed2021contrastive}
A.~Saeed, D.~Grangier, and N.~Zeghidour, ``Contrastive learning of
  general-purpose audio representations,'' in \emph{ICASSP 2021-2021 IEEE
  International Conference on Acoustics, Speech and Signal Processing
  (ICASSP)}.\hskip 1em plus 0.5em minus 0.4em\relax IEEE, 2021, pp. 3875--3879.

\bibitem{wang2022towards}
L.~Wang, P.~Luc, Y.~Wu, A.~Recasens, L.~Smaira, A.~Brock, A.~Jaegle, J.-B.
  Alayrac, S.~Dieleman, J.~Carreira \emph{et~al.}, ``Towards learning universal
  audio representations,'' in \emph{ICASSP 2022-2022 IEEE International
  Conference on Acoustics, Speech and Signal Processing (ICASSP)}.\hskip 1em
  plus 0.5em minus 0.4em\relax IEEE, 2022, pp. 4593--4597.

\bibitem{baevski2022data2vec}
A.~Baevski, W.-N. Hsu, Q.~Xu, A.~Babu, J.~Gu, and M.~Auli, ``Data2vec: A
  general framework for self-supervised learning in speech, vision and
  language,'' in \emph{International Conference on Machine Learning}.\hskip 1em
  plus 0.5em minus 0.4em\relax PMLR, 2022, pp. 1298--1312.

\bibitem{ragano2022learning}
A.~Ragano, E.~Benetos, and A.~Hines, ``Learning music representations with
  wav2vec 2.0,'' \emph{arXiv preprint arXiv:2210.15310}, 2022.

\bibitem{mccallum2022supervised}
M.~C. McCallum, F.~Korzeniowski, S.~Oramas, F.~Gouyon, and A.~F. Ehmann,
  ``Supervised and unsupervised learning of audio representations for music
  understanding,'' in \emph{ISMIR}, 2022.

\bibitem{grill2020bootstrap}
J.-B. Grill, F.~Strub, F.~Altch{\'e}, C.~Tallec, P.~Richemond, E.~Buchatskaya,
  C.~Doersch, B.~Avila~Pires, Z.~Guo, M.~Gheshlaghi~Azar \emph{et~al.},
  ``Bootstrap your own latent-a new approach to self-supervised learning,''
  \emph{Advances in neural information processing systems}, vol.~33, pp.
  21\,271--21\,284, 2020.

\bibitem{vaswani2017attention}
A.~Vaswani, N.~Shazeer, N.~Parmar, J.~Uszkoreit, L.~Jones, A.~N. Gomez,
  {\L}.~Kaiser, and I.~Polosukhin, ``Attention is all you need,''
  \emph{Advances in neural information processing systems}, vol.~30, 2017.

\bibitem{law2009evaluation}
E.~Law, K.~West, M.~I. Mandel, M.~Bay, and J.~S. Downie, ``Evaluation of
  algorithms using games: The case of music tagging.'' in \emph{ISMIR}.\hskip
  1em plus 0.5em minus 0.4em\relax Citeseer, 2009, pp. 387--392.

\bibitem{bogdanov2019mtg}
D.~Bogdanov, M.~Won, P.~Tovstogan, A.~Porter, and X.~Serra, ``The mtg-jamendo
  dataset for automatic music tagging.''\hskip 1em plus 0.5em minus 0.4em\relax
  ICML, 2019.

\bibitem{korzeniowski2017end}
F.~Korzeniowski and G.~Widmer, ``End-to-end musical key estimation using a
  convolutional neural network,'' in \emph{2017 25th European Signal Processing
  Conference (EUSIPCO)}.\hskip 1em plus 0.5em minus 0.4em\relax IEEE, 2017, pp.
  966--970.

\bibitem{knees2015two}
P.~Knees, {\'A}.~Faraldo~P{\'e}rez, H.~Boyer, R.~Vogl, S.~B{\"o}ck,
  F.~H{\"o}rschl{\"a}ger, M.~Le~Goff \emph{et~al.}, ``Two data sets for tempo
  estimation and key detection in electronic dance music annotated from user
  corrections,'' in \emph{Proceedings of the 16th International Society for
  Music Information Retrieval Conference (ISMIR); 2015 Oct 26-30; M{\'a}laga,
  Spain.[M{\'a}laga]: International Society for Music Information Retrieval,
  2015. p. 364-70.}\hskip 1em plus 0.5em minus 0.4em\relax International
  Society for Music Information Retrieval (ISMIR), 2015.

\bibitem{raffel2014mir_eval}
C.~Raffel, B.~McFee, E.~J. Humphrey, J.~Salamon, O.~Nieto, D.~Liang, D.~P.
  Ellis, and C.~C. Raffel, ``Mir\_eval: A transparent implementation of common
  mir metrics.'' in \emph{ISMIR}, 2014, pp. 367--372.

\bibitem{tzanetakis2002musical}
G.~Tzanetakis and P.~Cook, ``Musical genre classification of audio signals,''
  \emph{IEEE Transactions on speech and audio processing}, vol.~10, no.~5, pp.
  293--302, 2002.

\bibitem{kereliuk2015deep}
C.~Kereliuk, B.~L. Sturm, and J.~Larsen, ``Deep learning and music
  adversaries,'' \emph{IEEE Transactions on Multimedia}, vol.~17, no.~11, pp.
  2059--2071, 2015.

\bibitem{soleymani20131000}
M.~Soleymani, M.~N. Caro, E.~M. Schmidt, C.-Y. Sha, and Y.-H. Yang, ``1000
  songs for emotional analysis of music,'' in \emph{Proceedings of the 2nd ACM
  international workshop on Crowdsourcing for multimedia}, 2013, pp. 1--6.

\bibitem{engel2017neural}
J.~Engel, C.~Resnick, A.~Roberts, S.~Dieleman, M.~Norouzi, D.~Eck, and
  K.~Simonyan, ``Neural audio synthesis of musical notes with wavenet
  autoencoders,'' in \emph{International Conference on Machine Learning}.\hskip
  1em plus 0.5em minus 0.4em\relax PMLR, 2017, pp. 1068--1077.

\bibitem{wilkins2018vocalset}
J.~Wilkins, P.~Seetharaman, A.~Wahl, and B.~Pardo, ``Vocalset: A singing voice
  dataset.'' in \emph{ISMIR}, 2018, pp. 468--474.

\bibitem{yamamoto2022deformable}
Y.~Yamamoto, J.~Nam, and H.~Terasawa, ``Deformable cnn and imbalance-aware
  feature learning for singing technique classification,'' \emph{arXiv preprint
  arXiv:2206.12230}, 2022.

\bibitem{bock2016joint}
S.~B{\"o}ck, F.~Krebs, and G.~Widmer, ``Joint beat and downbeat tracking with
  recurrent neural networks.'' in \emph{ISMIR}.\hskip 1em plus 0.5em minus
  0.4em\relax New York City, 2016, pp. 255--261.

\bibitem{madmom}
S.~B{\"o}ck, F.~Korzeniowski, J.~Schl{\"u}ter, F.~Krebs, and G.~Widmer,
  ``{madmom: a new Python Audio and Music Signal Processing Library},'' in
  \emph{Proceedings of the 24th ACM International Conference on Multimedia},
  Amsterdam, The Netherlands, 10 2016, pp. 1174--1178.

\bibitem{marchand2015swing}
U.~Marchand and G.~Peeters, ``Swing ratio estimation,'' in \emph{Digital Audio
  Effects 2015 (Dafx15)}, 2015.

\bibitem{huang2020large}
Q.~Huang, A.~Jansen, L.~Zhang, D.~P. Ellis, R.~A. Saurous, and J.~Anderson,
  ``Large-scale weakly-supervised content embeddings for music recommendation
  and tagging,'' in \emph{ICASSP 2020-2020 IEEE International Conference on
  Acoustics, Speech and Signal Processing (ICASSP)}.\hskip 1em plus 0.5em minus
  0.4em\relax IEEE, 2020, pp. 8364--8368.

\bibitem{lee2018samplecnn}
J.~Lee, J.~Park, K.~L. Kim, and J.~Nam, ``Samplecnn: End-to-end deep
  convolutional neural networks using very small filters for music
  classification,'' \emph{Applied Sciences}, vol.~8, no.~1, p. 150, 2018.

\bibitem{modrzejewski2023transfer}
M.~Modrzejewski, P.~Szachewicz, and P.~Rokita, ``Transfer learning with deep
  neural embeddings for music classification tasks,'' in \emph{Artificial
  Intelligence and Soft Computing: 21st International Conference, ICAISC 2022,
  Zakopane, Poland, June 19--23, 2022, Proceedings, Part I}.\hskip 1em plus
  0.5em minus 0.4em\relax Springer, 2023, pp. 72--81.

\bibitem{heydari2021beatnet}
M.~Heydari, F.~Cwitkowitz, and Z.~Duan, ``Beatnet: Crnn and particle filtering
  for online joint beat downbeat and meter tracking,'' \emph{arXiv preprint
  arXiv:2108.03576}, 2021.

\bibitem{alonso2022music}
P.~Alonso-Jim{\'e}nez, X.~Serra, and D.~Bogdanov, ``Music representation
  learning based on editorial metadata from discogs,'' in \emph{ISMIR}, 2022.

\end{thebibliography}

\end{document}